\documentclass[sigconf,screen]{acmart}

\copyrightyear{2026}
\acmYear{2026}
\setcopyright{cc}
\setcctype{by}
\acmConference[CIKM '26]{Proceedings of the 35th ACM International Conference on Information and Knowledge Management}{November 07--11, 2026}{Rome, Italy}
\acmBooktitle{Proceedings of the 35th ACM International Conference on Information and Knowledge Management (CIKM '26), November 07--11, 2026, Rome, Italy}
\acmDOI{10.1145/3799682.3840250}
\acmISBN{979-8-4007-2539-5/2026/11}

\usepackage{soul}
\usepackage{xspace}
\usepackage{color, colortbl}  

\usepackage{adjustbox}
\usepackage{siunitx}
\usepackage{tabularx} 
\usepackage{amsmath}  
\usepackage{makecell}

\newcommand{\ourtool}{\textit{SearchLog}\xspace}

\usepackage{xcolor}

\definecolor{violet}{HTML}{6a51a3}
\definecolor{jade}{HTML}{00bb77}

\usepackage[english]{babel}  
\addto\extrasenglish{

}

\usepackage[capitalize]{cleveref} 
\usepackage[inline]{enumitem} 

\newcommand{\myparagraph}[1]{\paragraph*{\hspace*{-\parindent}\normalsize\bf#1}}
\AtBeginDocument{%
  }

\begin{document}

\title[SearchLog: A Web Browser Extension for Capturing Search Logs in Laboratory Studies]{SearchLog: A Web Browser Extension for Capturing Search Logs in Laboratory Studies}

\author{Jiaman He}
\orcid{0009-0007-2817-7675}
\affiliation{%
\institution{RMIT University}
\city{Naarm / Melbourne}
\country{Australia}
}
\email{jiaman.he@student.rmit.edu.au}

\author{Riccardo Xia}
\orcid{0009-0005-5274-7469}
\affiliation{%
\institution{Independent Researcher}
\city{Bologna}
\country{Italy}
}
\email{tcxia2001@gmail.com}

\author{Dana McKay}
\orcid{0000-0001-7522-1842}
\affiliation{%
\institution{RMIT University}
\city{Naarm / Melbourne}
\country{Australia}
}
\email{dana.mckay@rmit.edu.au}

\author{Damiano Spina}
\orcid{0000-0001-9913-433X}
\affiliation{%
\institution{RMIT University}
\city{Naarm / Melbourne}
\country{Australia}
}
\email{damiano.spina@rmit.edu.au}

\author{Johanne R. Trippas}
\orcid{0000-0002-7801-0239}
\affiliation{%
\institution{RMIT University}
\city{Naarm / Melbourne}
\country{Australia}
}
\email{j.trippas@rmit.edu.au}

\begin{abstract}

Natural search logs are valuable for studying search behavior in information seeking settings. We present \ourtool, an easy-to-install web browser extension for collecting natural search logs during lab-based studies. \ourtool enables participants to search the open web in a browser while recording structured interaction data from the mouse, keyboard, search activity, and browser state modules. 
The extension captures clicks, scrolling, hovered text, typed words, search queries, result rankings, AI-generated summaries when available, tab activity, and window changes. A local Flask backend stores each session as an ordered JSON event stream, including HTML snapshots and preprocessed search result data for later analysis. These logs can be used to derive measures such as query reformulation, page visits, dwell time, scroll behavior, tab switching, search path complexity, and exposure to AI-generated search content. By supporting natural browser-based search with structured experimental metadata, \ourtool provides a reusable resource to study search behavior across traditional and AI-enhanced search interfaces.
SearchLog is available at \url{https://github.com/peanutH/SearchLog-chromium-extension}.

\end{abstract}



\maketitle

\section{Introduction}
\label{sec:intro}
Understanding how and why people search requires studying their behaviors throughout the search process, not only the final answers they find. Search behaviors can reveal users' motivations, intentions, and decision-making during information seeking~\cite{Belkin2008challenge}. Prior work has shown that signals such as eye tracking, search logs, and physiological signals can help researchers understand users' knowledge, goals, personal characteristics, satisfaction, and search strategies~\cite{liu2016predicting,cole2015user,Liu2019satisfaction,he2026characterizing, he2025characterising, ji2024characterizing}. Among these signals, search logs have been widely used in studies of information-seeking behavior~\cite{silverstein1999analysis,jansen2000real,cole2015user,vuong2019naturalistic,liu2016predicting,zhang2025theory,liang2025flexible,he2026dc, trippas2024what, trippas2024reevaluating}. They offer an unobtrusive method for collecting large-scale search data from many system users, capturing interactions such as submitted queries, viewed results, clicks, and visited pages.

In laboratory studies, researchers may want participants to search naturally on the open web rather than use a custom search system~\cite{he2026dc, urgo2025search}. This is important because real-world search behavior is increasingly complex. Users may type and reformulate queries, view search result pages, open multiple pages, switch between tabs, scroll through content, return to earlier sources, and compare information across websites~\cite{granka2004eyetracking,joachims2005clickthrough,aula2010search,capra2009hcibrowser,capra2011hcibrowser,palani2025creative}. More recently, search engines have also begun to include AI-generated summaries, which introduce new ways for users to inspect, evaluate, and use search results~\cite{wardle2025evolving, allawati2026eye}. Capturing these actions is therefore important for studying how people search, evaluate, and use information in realistic online environments.

Collecting search logs can be challenging in practice. Existing approaches often use custom scripts, manual observation, screen recordings, browser-based research tools, or controlled search interfaces~\cite{capra2009hcibrowser,capra2011hcibrowser,makhlouf2021trackthinkts,palani2025creative}. These approaches can require extra setup, programming expertise, or manual review, and controlled interfaces may not fully reflect open web search behavior~\cite{vuong2019naturalistic,aula2010search}. General web analytics tools also mainly support website-level analysis, while researchers often need logs organized by participant, session, and task across websites and search engines.

This need has become more important because search result pages now show more than ranked links. They may also include AI-generated summaries, such as Google AI Overviews, that provide information directly on the results page. As a result, users may read and use information without opening external websites. Logs that only record queries, clicks, and visited pages may therefore miss important parts of the search experience

To support this data collection, we present \ourtool,\footnote{Code is available at \url{https://github.com/peanutH/SearchLog-chromium-extension}} an easy-to-install web extension that supports multiple Chromium-based browsers (e.g., Google Chrome, Opera, Brave, Microsoft Edge) for recording natural search behavior in laboratory studies. \ourtool lets participants search in a familiar browser while recording browser actions, page interactions, search result rankings, and, when available, an AI-generated summary. Currently, \ourtool supports two major commercial search engines (i.e., Google and Microsoft Bing) and detects their AI-generated summaries (i.e., AI Overviews, including conversations with AI Mode, and Copilot Search). Other search engines can be easily included using the same approach as described in \autoref{sec:Experiment} when needed as a future extension.

This paper makes the following contributions:

\begin{itemize}
    \item We introduce \ourtool, an easy-to-install web extension for collecting natural search logs during laboratory-based information-seeking experiments. A demo video is available.\footnote{Demo video is available at \url{https://youtu.be/nXHDAVfp7Zk}}

    \item We define a structured logging schema that captures browser state, browser events, page-level interactions, page information, search-result rankings, and AI-generated summaries across participants' search sessions.
\end{itemize}
\section{Related Work}
\label{sec:relatedwork}
\myparagraph{Interaction Logs in Information Seeking Research}

\begin{table*}[ht!]
    \centering
    \captionsetup{skip=5pt}
    \caption{Comparison of logging features across \ourtool and related exploratory search systems. 
    \checkmark indicates that the feature is explicitly logged or supported by the system.}
    \label{tab:logging_feature_comparison}
    \small
    \begin{tabular}{lccccccccc}
        \toprule
        \textbf{Tool} 
        & \makecell[c]{\textbf{Clicks}} 
        & \makecell[c]{\textbf{Scrolling}} 
        & \makecell[c]{\textbf{Mouse}\\\textbf{Movement}} 
        & \makecell[c]{\textbf{Hovered}\\\textbf{Text/Elements}} 
        & \makecell[c]{\textbf{Typed}\\\textbf{Text}} 
        & \makecell[c]{\textbf{Queries}}
        & \makecell[c]{\textbf{SERP}\\\textbf{Rankings}} 
        & \makecell[c]{\textbf{Tabs/}\\\textbf{Windows}}
        & \makecell[c]{\textbf{AI Summary}\\\textbf{Extraction}} \\
        
        \midrule
        
        SearchBar~\citep{morris2008searchbar} & --  & --  & --  & --  & --  & \checkmark  & --  & --  & -- \\
        
        TrackThinkTS~\citep{makhlouf2021privacy} & -- & \checkmark  & -- & -- & -- & \checkmark  & -- & \checkmark  & -- \\
        
        Search-Logger~\citep{singer2011search} & \checkmark  & --  & --  & --  & --  & \checkmark  & --  & \checkmark  & -- \\
        
        SearchPanel~\citep{qvarfordt2014searchpanel} & \checkmark  & --  & --  & --  & --  & \checkmark  & \checkmark  & --  & -- \\
        
        OrgBox~\citep{ward2021orgbox} & \checkmark  & --  & \checkmark  & \checkmark  & --  & \checkmark  & --  & --  & -- \\

        LogUI~\citep{maxwell2021logui} &  \checkmark  &  \checkmark  &  \checkmark  &  \checkmark  &  -- &  -- &  -- & \checkmark   & -- \\
        
        \textbf{\ourtool (Ours)} & \checkmark & \checkmark & \checkmark & \checkmark & \checkmark & \checkmark & \checkmark & \checkmark & \checkmark \\
        
        \bottomrule
    \end{tabular}
\end{table*}

Interaction logs are widely used to examine how users search, browse, compare options, evaluate information, and make decisions~\cite{wang2007learn, jansen2006we}. These logs are useful not only for interactive information retrieval and information-seeking research, but also for studies of consumer intent, online decision-making, and other web-based behavioral tasks~\cite{goel2010predicting, jansen2006search}. Common behavioral signals include submitted queries, clicked results, visited pages, dwell time, and scrolling behavior. These logs allow researchers to study the search process beyond final task outcomes, including how users formulate information needs, inspect results, reformulate queries, and compare information across sources~\citep{marchionini1995information, belkin1980anomalous, cole2015user, trippas2024reevaluating}.

In laboratory studies, however, collecting such logs in a natural search environment is not always convenient. Researchers may rely on custom logging scripts, screen recordings, browser histories, or manually coded observations. These approaches can be useful, but they often require technical expertise, substantial post-processing, or study-specific infrastructure. Controlled search interfaces can provide detailed logs, but they may also limit the naturalness of participants' searches on the open web. These limitations motivate the development of reusable tools that support structured logging while allowing participants to use familiar browser-based search environments.

\myparagraph{Browser-Based Logging Tools}

Several systems have been developed to support logging in web search and browsing studies. We have summarized some tools/systems in~\autoref{tab:logging_feature_comparison}. HCI Browser was designed to help researchers administer web search studies and collect browser event data, including visited websites, searches, windows, clicks, and scrolling events~\cite{capra2010hci}. YASBIL provides a browser-extension-based logging solution together with a WordPress plugin for capturing browsing activity during user studies~\cite{bhattacharya2021yasbil}. Other systems have supported search and browsing research through custom search panels, exploratory search interfaces, or controlled experimental platforms~\cite{qvarfordt2014searchpanel}.

These tools demonstrate the value of browser-based logging for information seeking research. However, many existing systems are tied to a specific interface, platform, or study workflow. In contrast, \ourtool is designed as an easy-to-install Chromium extension for laboratory studies in which researchers want participants to search naturally on the open web. Rather than replacing the participant's search environment with a custom interface, \ourtool records browser-level and page-level interactions during an explicitly started study session.

\myparagraph{Logging Search Result Pages with AI-generated Summaries}

Search result pages now show more than ranked links. They also include features such as AI-generated summaries that provide information directly on the page. As a result, users may read and use information on the results page without clicking external websites, making query-, click-, and URL-based logs insufficient to capture the full search process~\cite{li2009good,lagun2014attention,wu2020direct}.

\ourtool addresses this setting through a Chromium extension that is easy to install and collects richer browser-based search logs for laboratory studies. Compared with recent logging tools shown in~\autoref{tab:logging_feature_comparison}, \ourtool captures a broader set of events, including browser and page interactions, submitted queries, result rankings, and AI-generated summary contents when available. These logs are linked with participant, session, task, and timestamp metadata, enabling researchers to analyze behavior across both traditional and AI-enhanced search interfaces. Rather than introducing a new search interface, \ourtool provides a practical and reusable tool for collecting structured logs from natural web search.
\section{System Design and Logged Data}
\label{sec:Experiment}

\subsection{Architecture}
\label{subsec:architecture}

\begin{figure}[t]
    \centering
    \includegraphics[width=0.95\linewidth]{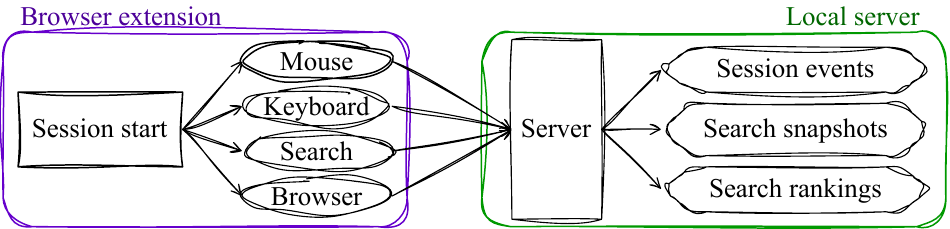}
    \caption{Workflow of \ourtool. After the researcher starts a session, participants naturally search in the web browser while the extension records their search and browser interactions. The local server saves the event logs and snapshots for later analysis.}
    \label{fig:searchlog-workflow}
\end{figure}

\ourtool is an operating system-independent toolkit with a client and a server component. The client consists of an extension for Chromium-based browsers that captures users' interactions. The server is responsible for tracking sessions and receiving and storing logged data. We depict the workflow \ourtool follows in~\autoref{fig:searchlog-workflow}.

\begin{figure}[t]
    \centering
    \includegraphics[width=0.6\linewidth]{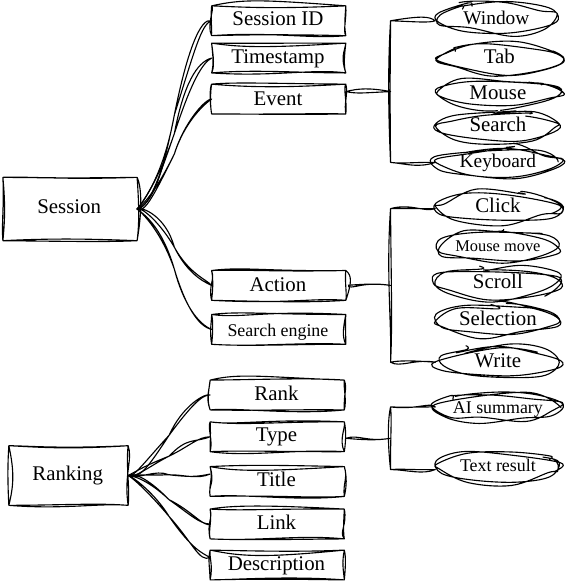}
    \caption{Structure of the data stored by \ourtool.}
    \label{fig:searchlog-structure}
\end{figure}

The extension records browser-level and page-level information. We organize the logged data into four modules, with each module capturing a different type of activity. Page-level data includes interactions with web pages, such as mouse actions, keyboard input, and search engine content. These data are collected using JavaScript event listeners and by analyzing the web page's Document Object Model (DOM) structure and parsing scripts injected into the page. Browser-level data includes actions related to browser windows and tabs, collected via the browser APIs. \autoref{fig:searchlog-structure} shows the data structure, and the code repository provides implementation details.

The local server is implemented in Flask and provides endpoints to start a session, receive logged data, and stop a session. When a session starts, the server creates a unique session ID and stores all received data in a dedicated local folder. 

Before a study session, the researcher installs the \ourtool Chromium extension and starts the local server on the experiment machine. The researcher then starts logging through the extension dialog, while participants use the browser naturally during the search task. After the task, the researcher stops logging through the extension dialog. Each session is saved as an ordered JSON event log, with search rankings and HTML snapshots stored separately. \autoref{fig:example_log} shows an example event log. Detailed installation and usage instructions are provided in the GitHub repository.

\begin{figure}
    \centering
\footnotesize
\begin{verbatim}
{
  "session_id": "1779009704_a1cb7ffe",
  "timestamp": 1779009711393,
  "event": "keyboard",
  "action": "write",
  "typed": "how solar panels work"
},
{
  "session_id": "1779009704_a1cb7ffe",
  "timestamp": 1779009712330,
  "event": "search",
  "search_engine": "google",
  "query": "how solar panels work",
  "filename_html": "1779009712330_how solar panels work.html",
  "filename_ranking": "1779009712330_how solar panels work.json"
},
{
  "session_id": "1779009704_a1cb7ffe",
  "timestamp": 1779009713110,
  "event": "mouse",
  "action": "mouse_move",
  "mouse_from": { "x": 923, "y": 386 },
  "mouse_to": { "x": 921, "y": 387 },
  "hovered_text": ""
}
\end{verbatim}
    \caption{Example of event logs produced by \ourtool.}
    \label{fig:example_log}
\end{figure}

\subsection{Session Lifecycle}
\label{subsec:lifecycle}

During a lab experiment, participants use a Chromium-based browser with \ourtool installed to complete search tasks. A session has four main steps.

\myparagraph{Session start.} The researcher starts the local server and begins logging through the extension dialog. The participant then starts the search task. Researchers should inform participants what data are collected and when logging starts and stops. In our setup, data are collected only during active study sessions and are stored locally.

\myparagraph{Logging.} During the task, participants search and browse as usual. \ourtool records their interactions in temporal order and saves events to disk as they occur, reducing the risk of data loss.

\myparagraph{Session end.} When the task ends, the researcher stops logging through the extension dialog. The system closes the session and removes injected code from the browser. A new session can then be started for the next participant.

\myparagraph{Artifacts for analysis.} Each session produces a local folder containing interaction logs, search result rankings, and page HTML snapshots. These files can be used to study behavior such as task time, page visits, dwell time, tab switching, scrolling, and query~events.

\ourtool is designed for researchers who need to collect natural search logs during laboratory-based information-seeking experiments. Instead of asking participants to use a custom search interface, \ourtool allows them to search on the open web in a Chromium-based browser. During the study session, the extension records browser and page interactions, including search result rankings and, when available, an AI-generated summary.

\myparagraph{Dry Run} To prevent logging errors or failures caused by interface changes or unexpected issues, we provide a dry-run function that simulates a search session. This allows users to verify that \ourtool is functioning correctly before beginning actual data collection.

\myparagraph{Supported Browsers and Search Engines} \ourtool is an extension for Chromium-based browsers, including Google Chrome, Microsoft Edge, Opera, and Brave. As search engines, we currently support Google and Bing, with the option to add others.

\myparagraph{Maintenance} By how \ourtool handles browser events and search engines, maintenance is required if the interfaces change. In particular, if the vendor of a search engine changes the layout of the page, \ourtool has to be updated accordingly to correctly extract data from the new layout.

\section{Technical Validation}
\label{sec:validation}

\myparagraph{Experiment Setup}

We conducted a small technical validation experiment using Google and Bing to examine whether \ourtool correctly records the main events generated during browser-based search sessions. The validation used non-sensitive demonstration tasks rather than real participant data. In each session, a tester started the local backend and browser extension, completed a predefined set of search actions, and then stopped the session.

The validation covered six test scenarios: (1) basic search, (2) multi-tab search, (3) mouse interaction, (4) keyboard input, (5) AI-generated summary capture, and (6) session-boundary handling. These scenarios were designed to cover common search behaviors, including submitting and reformulating queries, viewing search engine result pages, opening search results, scrolling web pages, hovering over text, typing in input fields, switching between tabs, and interacting with AI-generated summaries when available. We also included a session-boundary test to verify that logging stops after the session ends and that sensitive inputs, such as password fields, are masked.
The detailed validation scripts and the corresponding search logs are provided in the GitHub evaluation folder.

\myparagraph{Evaluation Criteria}

After each validation session, we inspected the generated log files to check whether \ourtool correctly recorded the expected events in temporal order. We verified that it created the required session files, captured browser, page, and search-specific interactions, logged AI-generated summaries when present, and stopped logging after the session ended. This evaluation tested the functionality and reliability of the logging pipeline. A real-time logging example is shown in our demo video.

\myparagraph{Results}

We observed that \ourtool always correctly initializes and terminates a session, properly creating the required files and directories on disk. Mouse and keyboard logs are correctly captured. Mouse movements are saved as start and end coordinates, each scroll is recorded, and hovered text is identified. Keyboard presses are recorded and grouped into a single entry if they occur within a certain time frame. Commands, such as selection and copy-paste, are also recognized. Input inserted in password fields is correctly hidden in the logs. When querying Google and Bing, a search event is correctly produced in the logs, and the HTML snapshot is saved. The parsed search rankings and AI-generated summary are also correctly saved to disk and are coherent with the page's content. Browser-level events involving tab and window operations are also correctly captured and recorded. We report a summary in \autoref{tab:validation-results}.

\begin{table}[t]
    \centering
    \caption{Summary of technical validation results.}
    \label{tab:validation-results}
    \small
    \begin{tabular}{l p{4.5cm}}
        \toprule
        \textbf{Component} & \textbf{Result} \\
        \midrule
        Session control & Starts and ends correctly. \\
        Clicks & Captures position. \\
        Scrolling & Captures start and end position. \\
        Mouse Movement & Captures start and end position. \\
        Hovered Text/Elements & Captures text content. \\
        Typed Text & Captures keys and commands. \\
        Queries & Detects search engine queries. \\
        SERP Rankings & Collects ranking correctly. \\
        Tabs/Windows & Captures open, close, focus, and loading. \\
        AI Summary Extraction & Detects content of AI summaries. \\
        HTML snapshots & HTML files saved correctly. \\
        \bottomrule
    \end{tabular}
\end{table}

\section{Final Remarks}
\label{sec:discussion}
\ourtool is released as a reusable research toolkit under Apache 2.0 license, including the Chromium extension, local Flask backend, documentation, and sample logs from non-sensitive demonstration tasks. Each session produces a local folder containing an ordered event stream, search result data, AI-generated summaries when available, and HTML snapshots. The documentation describes installation, session workflow, output structure, and example scripts for deriving common search behavior measures.

\myparagraph{Ethical Considerations} Because natural search logs may contain sensitive information, including URLs, page titles, typed text, search queries, AI-generated summaries, and page snapshots, \ourtool is designed for explicitly started laboratory sessions rather than continuous background monitoring. Researchers should clearly inform participants about what is collected, when logging starts and stops, how data are stored, and how withdrawal or deletion requests are handled. Study protocols should avoid personal accounts, passwords, or highly private search tasks, and researchers should collect only the data needed for their study.

\myparagraph{Limitations and Future Work} \ourtool currently has several limitations. It is designed for Chromium-based browsers and does not support Firefox and Safari without rewriting. In addition, search result extraction may require ongoing maintenance if commercial search engines update their page layouts. In future work, we plan to extend \ourtool to support more search engines (e.g., DuckDuckGo, Yahoo!, Baidu, Yandex, etc.) and to log interactions with LLM-based systems, such as ChatGPT, Copilot, Gemini, or Claude, using a similar approach that analyzes the page DOM and extracts relevant interaction content.

\begin{acks}

This research is supported by \grantsponsor{ARC}{Australian Research Council (ARC)}{https://www.arc.gov.au/} Centre of Excellence for Automated Decision-Making and Society (ADM+S, \grantnum{ARC}{CE200100005}).

\end{acks}

\subsection*{GenAI Usage Disclosure}

During the preparation of this work, the author used ChatGPT in order to assist with language editing and improving the clarity of portions of the manuscript. The AI tools were not used to generate data, conduct analyses, or make scientific decisions. After using this tool/service, the author reviewed and edited the content as needed and take full responsibility for the content of the published article.

\bibliographystyle{ACM-Reference-Format}
\balance
\bibliography{00-refs}

\end{document}